\documentclass[aps,preprint]{revtex4}
\usepackage{graphicx}

\draft
\begin{document}

\title{Lithium Ionization by a Strong Laser Field}
\author{ Camilo Ruiz, Luis Plaja, and Luis Roso}
\address{Departamento de F\'\i sica Aplicada, Universidad
de Salamanca, E-37008 Salamanca, Spain}
\date{\today}

\begin{abstract}
We study {\it ab initio} computations of the interaction of Lithium with a strong laser field. Numerical solutions of the time-dependent fully-correlated three-particle Schr\"odinger equation restricted to the  one-dimensional soft-core approximation are presented. Our results show a clear transition from non-sequential to sequential double ionization for increasing intensities. Non sequential double ionization is found to be sensitive to the spin configuration of the ionized pair. This asymmetry, also found in experiments of photoionization of Li with synchrotron radiation, shows the evidence of the influence of the exclusion principle in the underlying  rescattering mechanism.
\end{abstract}


\maketitle

\newpage

Photoionization of atoms by short-pulsed intense laser radiation constitutes an extraordinary playground to test quantum mechanics beyond the perturbative limit. The earliest experiments on this subject already showed many new features unexplained by the standard approaches of that time, and renewed the interest on alternative non-perturbative theoretical developments. Among them, the {\em ab initio} numerical integration  of the Schr\"odinger equation provided a fundamental tool for the profound understanding of the dynamics of laser-matter interaction. Always limited by the current state-of-the-art of the computing hardware, the numerical codes first   targeted the problem of single electron ionization in one dimension using a soft-core potential \cite{Eberly0,Eberly1,Eberly2,Eberly3,Eberly4}. The reduction to 1D has been considered for more than ten years a reasonable strategy to get a first insight into the dynamics of intense field ionization.  In particular, many aspects of the strong field phenomena can be described qualitatively at this level (among them, above threshold ionization, and the basic general features of the harmonic spectrum). More recently,  one-electron three-dimensional codes have helped to unravel more subtile phenomena as non-dipolar effects \cite{javier3D}. However, 
the complexity of the numerical task grows exponentially with the number of particles, as new dimensions have to be added. This introduces a rather serious technical limit to the computation of the full 3D  dynamics of more than one particle. Nowadays, the exact integration of the 3D Schr\"odinger equation can be accomplished only for the case of He interacting with linearly polarized electromagnetic fields, employing a extraordinary amount of computing resources \cite{Taylor2000,Taylor2001}. In these circumstances, the dimensional reduction of the many-particle problem continues as a fundamental tool. For instance, the 1D approach to the laser-He interaction \cite{Eberly4,Gross} is still employed as the most common technique to tackle the two-particle problem. For three electron problem, the dimensional reduction appears almost mandatory. Note that other traditional approaches (density functional theories) have not straightforward applications in the limit of small number of particles  \cite{Valerie}, which are highly correlated. Quantum correlations, therefore, play a fundamental role in the dynamics of few particles. The advantage of the dimensional reduction is to allow {\em ab initio} numerical calculations that include completely these correlations.

Accordingly to the underlying mechanism, the double photoionization of Helium can be cataloged as  sequential or non sequential. In the first case, both electrons ionized independently by photon absorption from the electromagnetic field. In contrast, the non-sequential ionization reveals a more subtile dynamics, in which the second electron is ionized via scattering with the first \cite{ABeckerReview,ABecker2003,WBecker}.  One signature of the relevance of quantum correlation in this later process consists in the sensitivity to the particular form of entanglement of the ionizing pair. As reported in \cite{Salamanca1,Salamanca2,Gibson1}, the rescattering process is less effective when the two-electron wavefunction is antisymmetrized in the orbital part (orthoHelium) rather than in the spin part (paraHelium). In the three particle problem, i.e. Lithium, correlations appear more intrincated involving non-separable orbital and spin antisymmetries.   Experimental  works on the double and triple ionization in Lithium has been published recently for synchrotron radiation and ion or electron collision  \cite{ Huang, Wehlitz2002,Wehlitz2004,Nico}.  On the other hand,  previous theoretical treatments include the high photon energy limit \cite{derHart}, approximated half-collision models \cite{Pattard1,Pattard2} and, very recently, close-coupling  grid calculations in the weak field limit  \cite{colgan}.

To our knowledge, the problem of Li photoionization in strong laser fields using {\em ab initio} numerical calculations of the three-particle problem has not been previously addressed. Of course, the full 3D problem falls well beyond present and near future computing capabilities. However, the problem in reduced dimensionality (1D for each particle) can be addressed with a medium-size computer. We, therefore, present in this paper the first results of this type of calculations that consider fully correlated electrons. 

In the limit of very high photon energies, "shake off" has been determined to be the main mechanism for double and triple ionization of Lithium  \cite{derHart,Huang}. However, in the case of photon energies below some hundreds of eV, a different mechanism has been proposed \cite{Samson, Pattard1, Nico}. In this case the electromagnetic field ionizes one or two electrons from the inner K-shell which, in the way out,  ionize one of the remaining electrons. This viewpoint seems to be confirmed experimentally by Wehlitz {\it et al} \cite{Wehlitz2002}, with synchrotron radiation. In addition, the same experiment suggests through the comparison with photoionization of He, that the double ionization of Li is not equally efficient for the different spin configurations of the ionized pair of electrons.  The present study confirms this aspect and gives a fundamental description in terms of the inhibition of $e^- \rightarrow 2e^-$ scattering due to the Pauli's exclusion principle.   As stated previously \cite{Salamanca1, Gibson1}, the  symmetric character of the spatial wavefunction with respect to exchange of particles, can inhibit non-sequential double ionization.

We construct the three-particle hamiltonian in reduced dimensionality  by extension of the previous models for Hydrogen and Helium atoms (in a.u.):
\begin{equation}
\label{hamiltonian}
 H_0=\sum^3_{i=1}\left( {p_i^2\over 2}-{3\over \sqrt{a^2+z^2_i}}\right)+\sum_{i\neq j}{1\over \sqrt{b^2+(z_i-z_j)^2}}
\end{equation}
where $a$ and $b$ are the parameters of the soft potential. This form of hamiltonian commutes with the symmetry operators and, therefore, the symmetry of the wavefunction remains as a constant of motion. Initially, we will assume the atom in its ground state  $^2S_{1/2}$, therefore the wavefunction at any time has this symmetry and may be expressed  as
\begin{eqnarray}
\label{eq:ModelWave}
\Phi_{\alpha \alpha \beta} (z_1,z_2,z_3, t)&\propto&\alpha(1)\alpha(2)\beta(3)\phi_{12}(z_1,z_2,z_3,t)\nonumber \\
&+&\alpha(1)\beta(2)\alpha(3)\phi_{13}(z_1,z_2,z_3,t) \nonumber \\
&+&\beta(1)\alpha(2)\alpha(3)\phi_{23}(z_1,z_2,z_3,t)
\end{eqnarray}
The spin part is the combination of three single electron spin functions, in our case $\alpha(i) \equiv | \downarrow \rangle $ and $\beta(i)\equiv | \uparrow  \rangle$. The orbital functions $\phi_{ij}(z_1,z_2,z_3,t)$ are antisymmetric under the permutation $i \leftrightarrow j$. Note that we have  written (\ref{eq:ModelWave}) in such a way that the different terms in the summation have orthogonal spin states. This form is particularly useful with the non-relativistic hamiltonian (\ref{hamiltonian}) since the spin state is a constant of motion and, therefore, every term in the summation evolves independently from the others. Moreover, it will be only necessary to compute the time evolution of one of them, since the others can be found by simple permutations. 

The  ground state of our model hamiltonian is computed using imaginary-time evolution with an initial trial function for  $\phi_{ij}(z_1,z_2,z_3,t=0)$ with the required symmetry. The soft core potentials parameters were used to fit the energy of the ground state to the experimental value $E=-7.33$ a.u. (199.44 eV), i.e. $a=b=0.4969$ (0.262 $\AA$) 
Once the ground state with the required accuracy is found, we propagate it in time according to the minimal coupling Hamiltonian:
\begin{equation}
{i\partial \over \partial t} \Phi_{\alpha \alpha \beta} (z_1,z_2,z_3, t)=\left[ H_0+(p_1+p_2+p_3)  A(t)/c \right] \Phi_{\alpha \alpha \beta} (z_1,z_2,z_3, t)
\end{equation}
The vector potential $A(t)$ is assumed to be linearly polarized along the dimension described in the model (as usual, the $A^2(t)$ term of the interaction hamiltonian has been factorized as a global phase). As is standard in the Helium case \cite{Eberly4}, the ionization yield is computed using a partition of the Hilbert space. The extension to the Lithium case reads as:
\begin{equation}
\label{eq:espreg}
\cases{ Li &if $|z_1|<15$ a.u. and $|z_2|<15$ a.u. and $|z_3|<15$ a.u. \cr Li^+ &if $|z_i|>15$ a.u. and $|z_j|<15$ a.u. and $|z_k|<15$ a.u. \cr Li^{2+} &if $|z_i|>15$ a.u. and $|z_j|>15$ a.u. and $|z_k|<15$ a .u.  \cr Li^{3+} &elsewhere }
\end{equation}
with $i,j,k=1,2,3$

The total ionization yield is obtained by adding  the contributions of each of the three terms in eq.  (\ref{eq:ModelWave}), which describe orthogonal spin configurations. Inspired by the synchrotron experiments, we present calculations of the ionization of one-dimensional Li with a pulse of frequency of $\omega=3.016$ a.u. (82.06 eV) and intensities ranging from $I=10^{-3}$ a.u. ($3.5\times 10^{13}$ W/cm$^2$) up to $I=10$ a.u. ($3.5\times 10^{17}$ W/cm$^2$). High power coherent radiation in this wavelengths are expected to be available  at the end of this year in phase 2 of the FEL- TTF at Hasylab (Hamburg). To achieve the relevant intensities used in this paper, a slight focussing would be needed to focal spots of the order of $10 \mu m$. The length of the pulse duration (four cycles) is limited by our computer's capabilities. Larger pulse durations are expected to increase the ionization yield, but not to alter the fundamental mechanism. Note that, specially in the case of shorter pulses, the computations have to be carried out over a time interval large enough to allow the ionized population to drift into the proper spatial regions (\ref{eq:espreg}). This interval is typically larger than the pulse length and is determined according to the saturation of the ionization yields (see for instance Fig. \ref{fig:yields}a).  

Figure \ref{fig:ratio} shows the ratio of double to single ionization yields for different intensities, computed at about eight laser periods after the end of the interaction. As well established in the photoionization of Helium, the change in the slope of this ratio as the intensity increases (often referred as knee) is the signature of the transition from non-sequential to sequential double ionization. Hence, this figure demonstrates, that double ionization of Lithium also shifts from non-sequential to sequential as the intensity increases. In our particular case, we may take $I_{th}=10^{15}$  W/cm$^2$ as the threshold value between these two mechanisms. However, in contrast with the Helium case, in Lithium there are two possible channels of correlated double ionization. They correspond to the two different spin configurations of the ionized pair:  parallel $(\alpha\alpha)$  or antiparallel $(\alpha\beta)$. Note that  the wavefunctions $\phi_{ij}(z_1,z_2,z_3)$ correspond to a definite spin orientation in every coordinate. Therefore, a further partition of the spatial volume corresponding to double ionization permits us to  track this two channels separately.  For instance,  in the particular case of $\phi_{12}(z_1,z_2,z_3)$ , the volumes $|z_1|<15$ a.u., $|z_2|>15$ a.u.,  $|z_3|>15$ a.u.,  and  $|z_1|>15$ a.u., $|z_2|<15$ a.u.,  $|z_3|>15$ a.u. describe double ionization of an entangled pair with antiparallel spins, while  $|z_1|>15$ a.u., $|z_2|>15$ a.u.,  $|z_3|<15$ a.u. describes the parallel configuration. Figure \ref{fig:yields}a shows the dynamics of double ionization in each of these two channels, as a function of time at different laser intensities. As noted before, the ionized population takes some time to access  the spatial regions where it is computed. This can be seen in the figure as the ionization yield stabilizes at times larger than the interaction. It becomes also apparent the different dynamics of ionization for each spin configuration at intensities below the threshold $I_{th}$, i.e. when non-sequential ionization is the relevant mechanism. On the contrary, both channels tend to be equally possible when  the ionization is sequential. In conclusion, ionization of electron pairs with antiparallel configuration is shown to be more probable when non-sequential ionization takes place. Figure \ref{fig:yields}b plots the relative difference of the ionization yields at the end of the computation (final points in Fig.  \ref{fig:yields}a), which is tipically above $50\%$ in the non sequential case. This result is in clear agreement with the indication in \cite{Wehlitz2002}, in the sense that comparison of their experimental results with the ionization of Helium would suggest such asymmetry. On the other hand, our previous work in ionization of Helium has shown that the $e^- \rightarrow 2e^-$ scattering process is greatly inhibited for the orthohelium case, since the parallel spin configuration implies the antisymmetric character of the orbital wavefunction, in which Pauli's principle reduces the strength of electron-electron interaction \cite{Salamanca1}. Figure \ref{fig:density} demonstrates that this also the case in the double ionization in Li. It shows the density distribution corresponding to the term $\alpha(1)\alpha(2)\beta(3)\phi_{12}(z_1,z_2,z_3,t)$ in (\ref{eq:ModelWave}) at the end of the computation at the planes $z_1=0$, $z_2=0$ and $z_3=0$. To improve legibility, black lines outline the limits between the regions corresponding to the neutral Li, Li$^+$ and Li$^{2+}$. As discussed previously,  the double ionization is represented by the out-of-axis regions. In this particular case, the vertical planes correspond to double ionization of a electron pair with antiparallel spins, while the horizontal corresponds to the parallel configuration. The inhibition of this later case is, therefore, apparent from this plot. Therefore, and in agreement with  \cite{Wehlitz2002}, the dominant mechanism of non-sequential double ionization of Lithium at these frequencies consists in a first release of an electron, followed almost instantaneously by the scattering with one of the two remaining electrons. The exclusion principle makes this scattering most effective for the antiparallel spin configuration, hence resulting in a larger ionization yield. 

In conclusion, we have presented {\em ab initio} results for the interaction of Lithium with a strong laser field, in a reduced geometry. The model, that has been proven to give deep qualitative insight on this process for the simplest atoms (Hydrogen and Helium), is developed taking into account the three interacting electrons on equal footing with the proper symmetrization of the wavefunction, and including full account of quantum correlations. Our results reveal the asymmetry of the non-sequential double ionization process in relation with the spin configuration of the entangled ionized pair. We give fundamental insights of this phenomena, based in the sensitivity of the electron rescattering to the symmetry of the orbital wavefunction.

\section{ACNOWLEDGEMENTS}
This work has been partially supported by the Spanish
Ministerio de Ciencia y Tecnologia (FEDER funds, grant
BFM2002-00033)
and by the Junta de Castilla y Le\'on (grant SA107/03).

\newpage

\begin{figure}
\caption{Double to single ionization ratio. By summing over all spin configurations we can obtain the total ionization yield as a function of the intensity. The "knee" structure below the $I=10^{15}$ W/cm$^2$, has been recognized as an indicator of the correlated nature of the ionization process.}
\label{fig:ratio}
\end{figure}

\begin{figure}
\caption{ (a) Double ionization on each of the two possible channels as a function of time. Solid line represents the channel with parallel spins and dashed line represents the channel with antiparallel spins. For low intensities the yields are different because of the inhibition mechanism. (b) Relative difference for each of the channels as a function of the intensity. Channel 1 corresponds to ionization with parallel spin $| \uparrow\uparrow\rangle$ electrons ionized in the final state, and channel 2 correspond to antiparallel spin $| \uparrow\downarrow\rangle$ electrons ionized in the final state.}
\label{fig:yields}
\end{figure}

\begin{figure}
\caption{ Slices of the three dimensional density of the function $|\phi_{12}(z_1,z_2,z_3,t)|^2$ in logarithmic scale, corresponding to the spin configuration $\alpha(1)\alpha(2)\beta(3)$,  at the end of the pulse for $I=10^{13}$ W/cm$^2$. Population in the vertical planes outside the fringes at 15 a.u. (7.9 $\AA$ ) corresponds to double ionization with antiparallel spin electrons  ($|\uparrow\downarrow\rangle$ ). The population in the horizontal plane outside the fringes at 15 a.u. corresponds to double ionization with parallel spin electrons   ($|\uparrow\uparrow\rangle$ ). Pauli principle inhibits double ionization at the $z_1=z_2$ plane, because of the antisymmetry of the wavefunction. }
\label{fig:density}
\end{figure}


\begin{references}


\bibitem{Eberly0} R. Grobe and J. H. Eberly, Phys. Rev. Lett. {\bf 68} 2905 (1992) 
\bibitem{Eberly1} S L Haan, R Grobe, and J H Eberly, Phys. Rev. A {\bf 50} 378 (1994) 
\bibitem{Eberly2} Q. Su, J.H. Eberly Phys. Rev. A {\bf 44} 5997 (1991) 
\bibitem{Eberly3} S. L. Haan, P. S. Wheeler, R. Panfili and J. H. Eberly. Phys. Rev. A.  {\bf 66}, 061402(R) (2002).
\bibitem{Eberly4} C. Szymanowski, R. Panfili,W.-C. Liu, S. L. Haan and J. H. Eberly, Phys. Rev A. {\bf 61} 055401 (2000)
\bibitem{Gross} M. Lein, E. K. U. Gross, and V. Engel, Phys. Rev. Lett, {\bf 85}, 4707, (2000).

\bibitem{javier3D}  N.J. Kylstra, R.A. Worthington, A. Patel,  P.L. Knight,  J.R. V\'azquez de Aldana, L. Roso Phys. Rev. Lett. {\bf 85} 1835 (2000)

\bibitem{Taylor2000} Jonathan S Parker, Laura R Moore, Daniel Dundas and K T Taylor, J. Phys. B. {\bf 33} L691 (2000)
\bibitem{Taylor2001} J S Parker, L R Moore, K J Meharg, D Dundas and K T Taylor. J. Phys. B. {\bf 34} L69 (2001) 

\bibitem{Valerie} V. Veniard, R. Taieb, and A. Maquet, Laser Phys.  {\bf 13}, 465, (2003)

\bibitem{ABeckerReview} A. Becker and F. H. Faisal,ÊOpt. Express {\bf 8}, 383-394 (2001)
\bibitem{ABecker2003} A. Becker and F. H.M. Faisal, Phys. Rev. Lett. {\bf 89} 193003 (2002)
\bibitem {WBecker} R. Kopold, W. Becker, H. Rottke, and W. Sandner. Phys. Rev. Lett. {\bf 85}, 3781 (2000)

\bibitem{Salamanca1} Camilo Ruiz, Luis Plaja, J. R V\'azquez de Aldana and Luis Roso. Phys. Rev. A. {\bf 68}, 023409-1, (2003)
\bibitem{Salamanca2} C. Ruiz, L. Plaja, J. R. V\'azquez de Aldana and L. Roso. App. Phys. B. {\bf 78}, 829, (2004).
\bibitem{Gibson1} C. Guo, R. T. Jones, and G. N. Gibson, Phys. Rev. A {\bf 62}, 015402 (2000).


\bibitem{Wehlitz2002} R. Wehlitz and J. B. Bluett, S. B. Whitfield Phys. Rev. A  {\bf  66}, 012701 (2002)
\bibitem{Wehlitz2004} R. Wehlitz, M. M. Martinez, J. B. Bluett, D. Lukic, and S. B. Whitfield. Phys. Rev. A {\bf 69}, 062709. (2004)
\bibitem{Huang} M.-T. Huang, W. W. Wong, M. Inokuti, S. H. Southworth and L. Young, Phys. Rev. Lett. {\bf 90}, 163201-1, (2003).
\bibitem{Nico} J.A. Tanis, J.Y. Chesnel, F. Fr\'emont, D. Hennecart, X. Husson, A. Cassimi, J.P. Grandin, B. Skogvall, B. Sulik, J.-H. Bremer and N. Stolterfoht. Phys. Rev. Lett. {\bf 83}, 1131, (1999).


\bibitem{Kheifets} A. S. Kheifets, A. Ipatov,  M. Arifin and Igor Bray. Phys Rev A. {\bf 62} 052724 (2004).
\bibitem{derHart} Hugo W. van der Hart and Chris H. Greene, Phys. Rev. Lett.{\bf 81}, 4333, (1998)


\bibitem{Samson} James A. R. Samson. Phys. Rev. Lett. {\bf 65}, 2861, (1990).
\bibitem{Pattard1} Thomas Pattard and Joachim Burgd\"orfer, Phys. Rev. A. {\bf 64}, 042720, (2001).
\bibitem{Pattard2} Thomas Pattard and Joachim Burgd\"orfer, Phys. Rev. A. {\bf 63}, 020701, (2001).
\bibitem{colgan} J. Colgan, M. S. Pindzola and F. Robicheaux, Phys. Rev. Lett. {\bf 93} 053201 (2004)


\end{references}
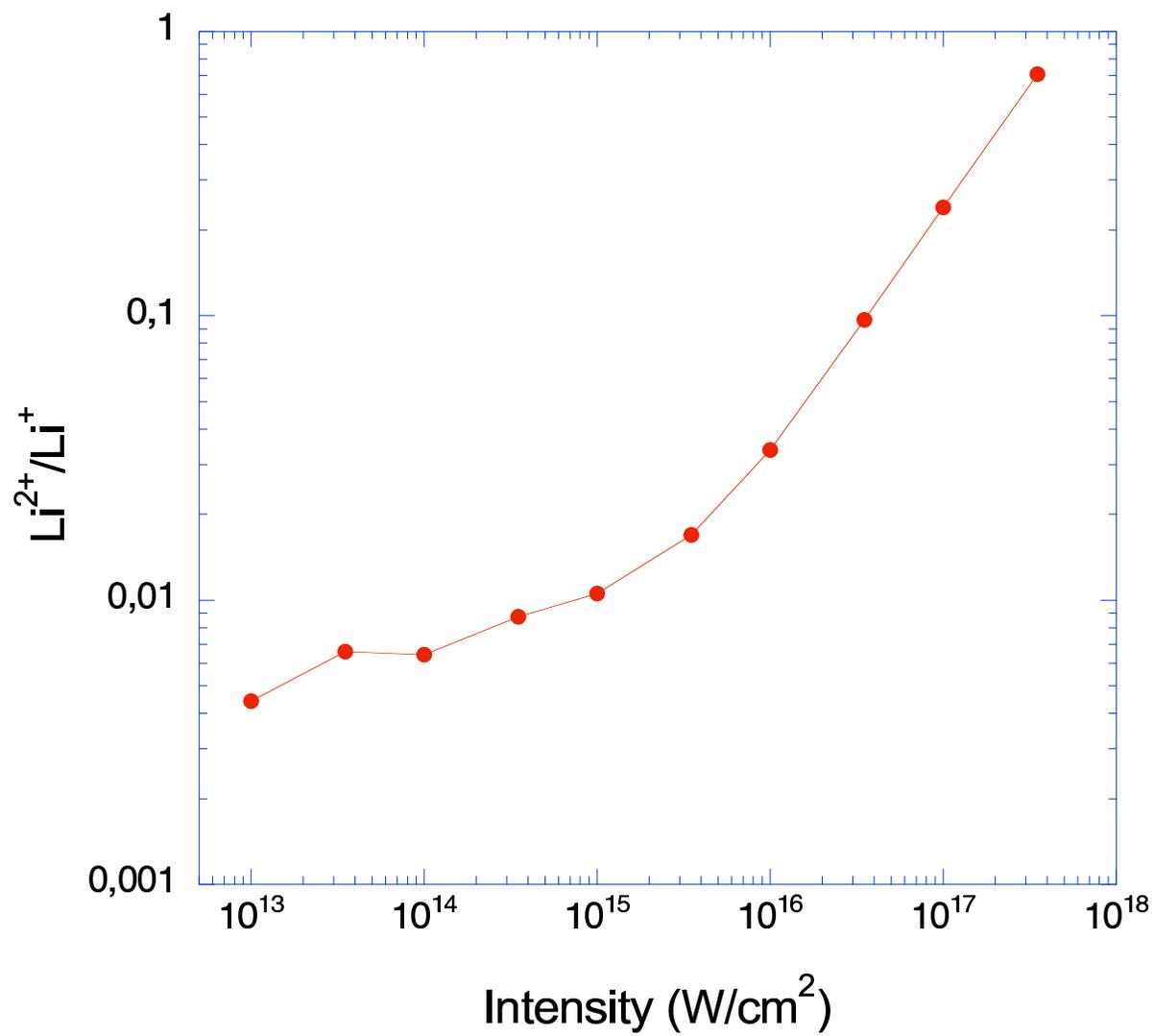
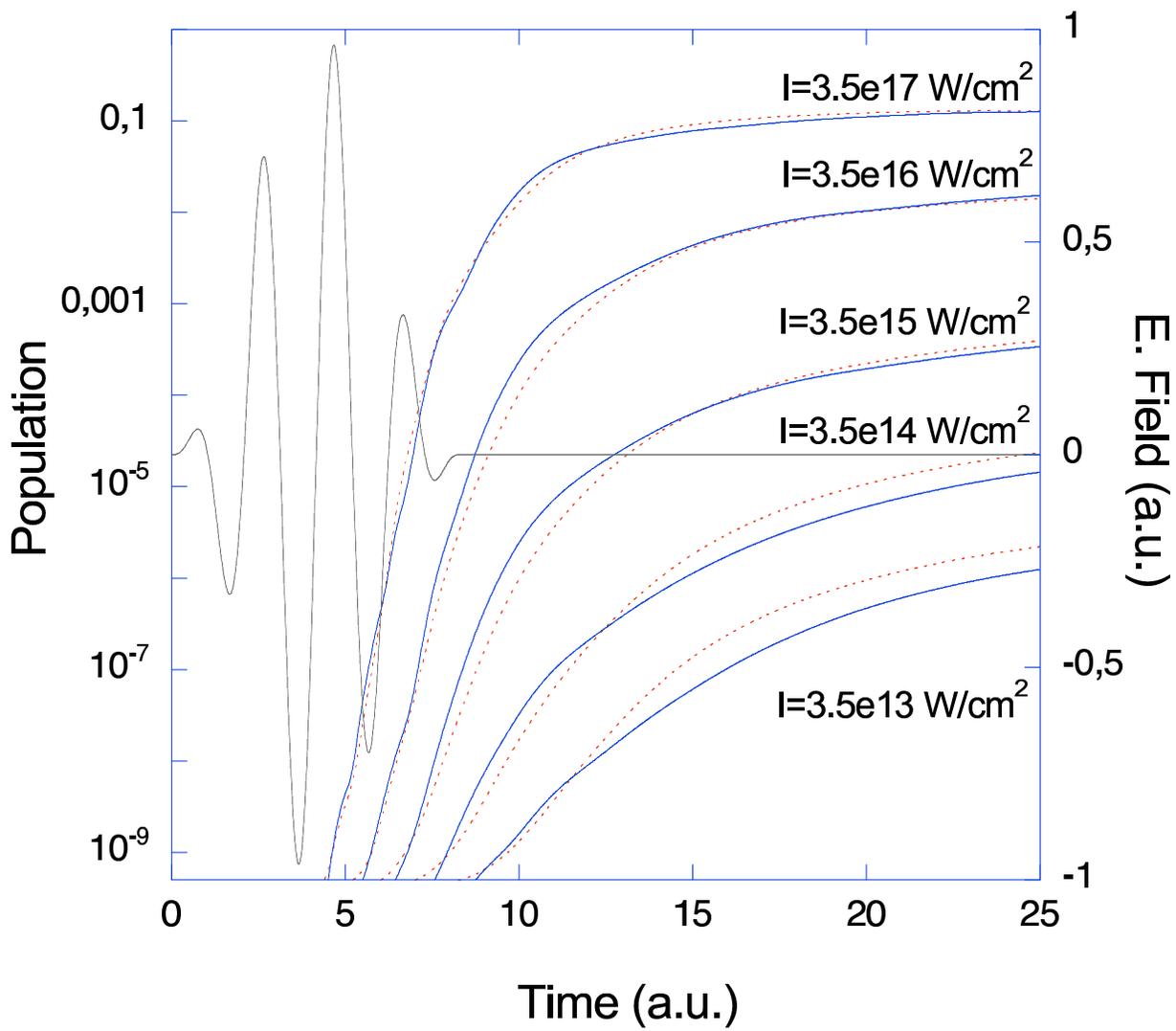
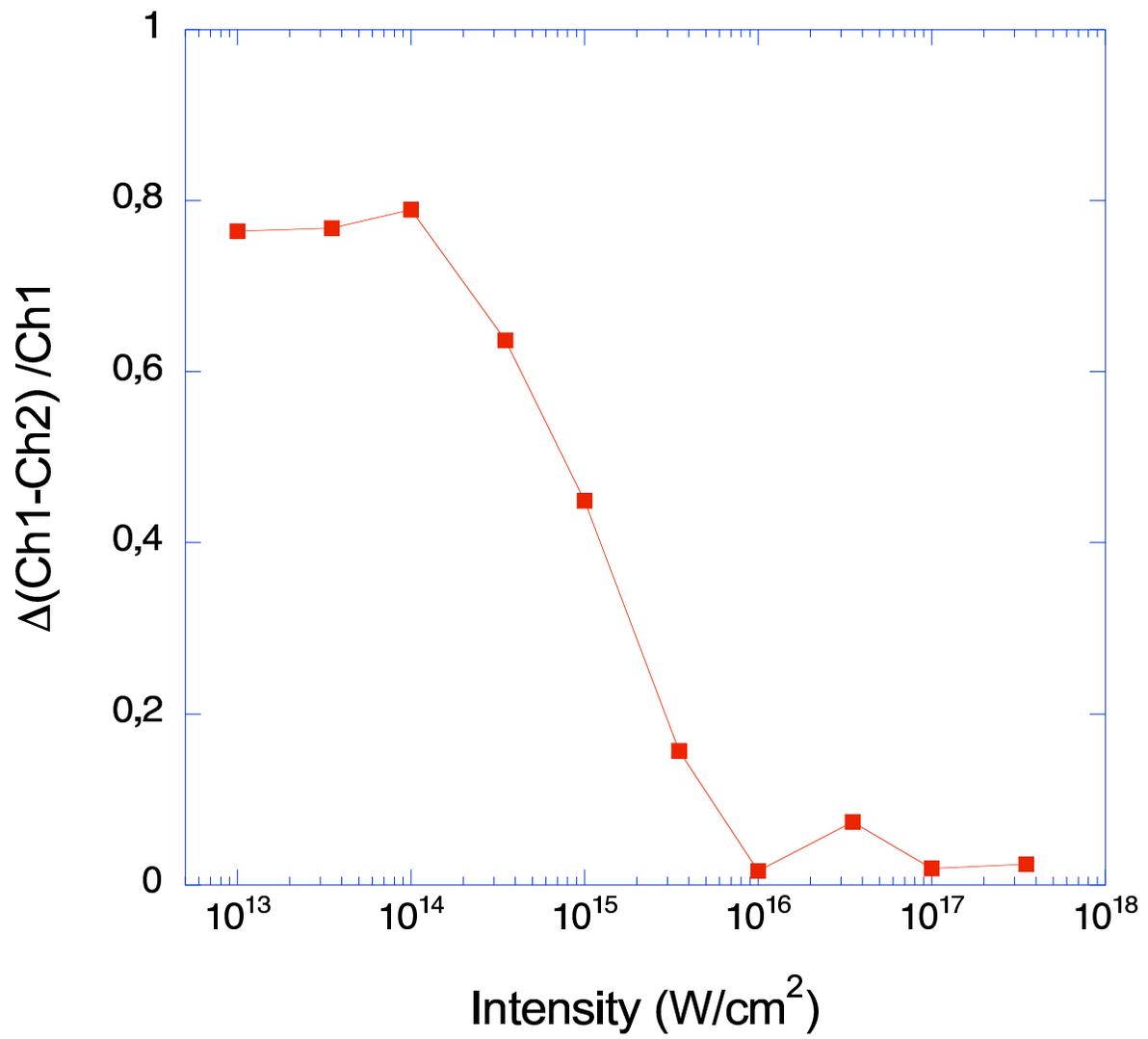
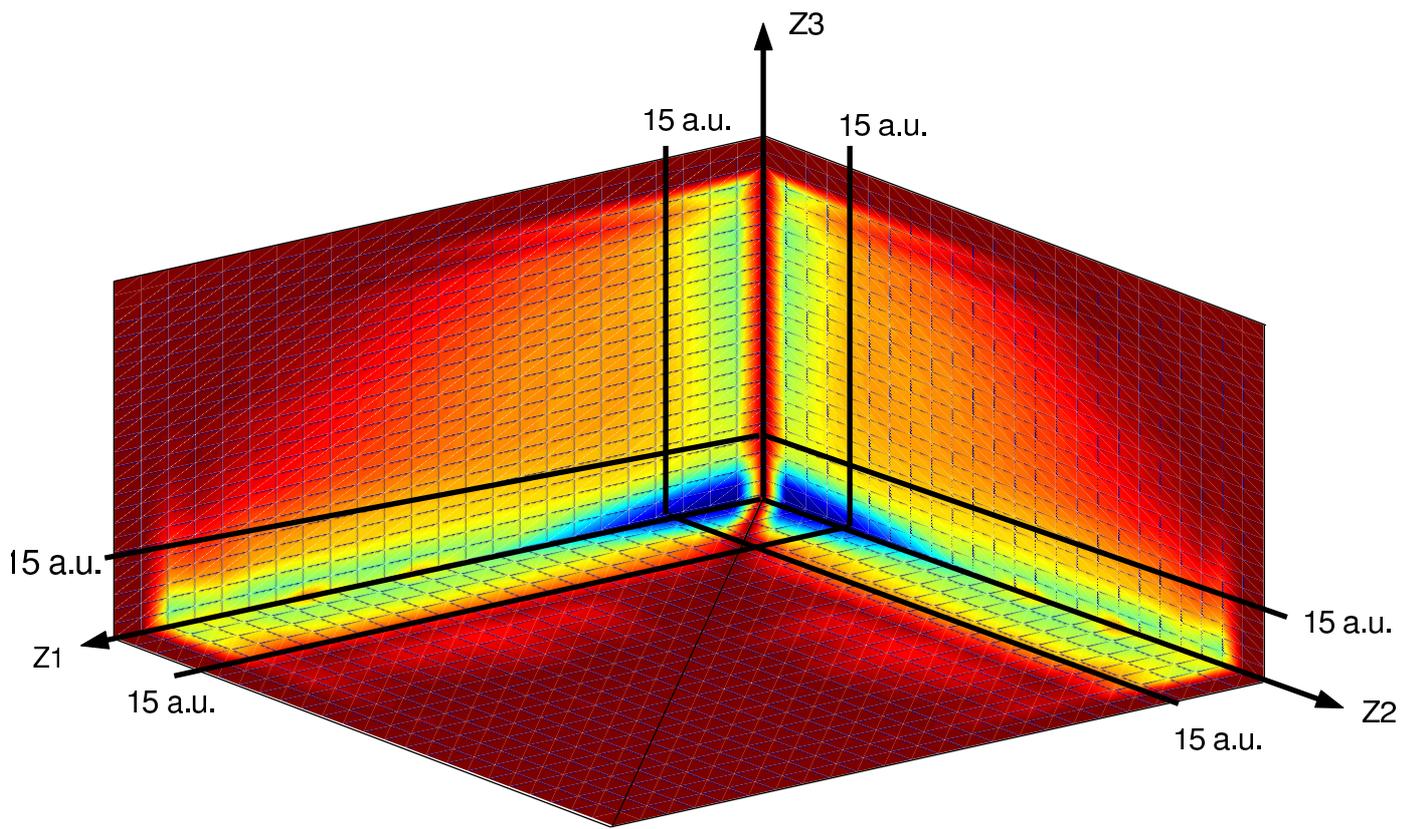
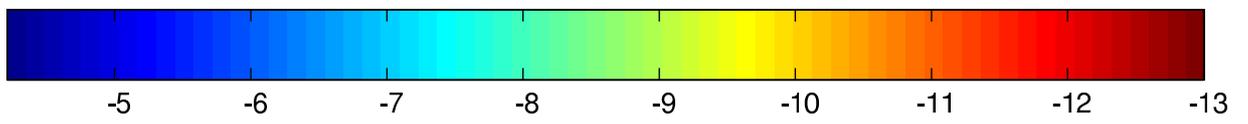
\end{document}